\documentclass{emulateapj}

\usepackage{graphicx}
\usepackage{amssymb}
\usepackage{amsmath}
\usepackage{times}

\newcommand{\lsi}{LS~I~+61$^{\circ}$303}
\newcommand{\hess}{HESS~J0632+057}
\newcommand{\be}{MWC~148}

\begin{document}
\title{Detection of VHE $\gamma$-rays from \hess\ during the 2011 February  X-ray outburst with the MAGIC Telescopes }
\shorttitle{VHE $\gamma$-ray emission from \hess\ }
\shortauthors{Aleksi\'c et~al.}
\author{
J.~Aleksi\'c\altaffilmark{1},
E.~A.~Alvarez\altaffilmark{2},
L.~A.~Antonelli\altaffilmark{3},
P.~Antoranz\altaffilmark{4},
M.~Asensio\altaffilmark{2},
M.~Backes\altaffilmark{5},
U.~Barres de Almeida\altaffilmark{6},
J.~A.~Barrio\altaffilmark{2},
D.~Bastieri\altaffilmark{7},
J.~Becerra Gonz\'alez\altaffilmark{8}
% ...
}
\altaffiltext{1} {IFAE, Edifici Cn., Campus UAB, E-08193 Bellaterra, Spain}
\altaffiltext{2} {Universidad Complutense, E-28040 Madrid, Spain}
\altaffiltext{3} {INAF National Institute for Astrophysics, I-00136 Rome, Italy}
\altaffiltext{4} {Universit\`a  di Siena, and INFN Pisa, I-53100 Siena, Italy}
\altaffiltext{5} {Technische Universit\"at Dortmund, D-44221 Dortmund, Germany}
\altaffiltext{6} {Max-Planck-Institut f\"ur Physik, D-80805 M\"unchen, Germany}
\altaffiltext{7} {Universit\`a di Padova and INFN, I-35131 Padova, Italy}
\altaffiltext{8} {Inst. de Astrof\'{\i}sica de Canarias, E-38200 La Laguna, Tenerife, Spain}
% ...

\begin{abstract}

The very high energy (VHE) $\gamma$-ray source \hess\ has recently been confirmed to be a $\gamma$-ray binary. The optical counterpart is the Be star \be, and a compact object of unknown nature orbits it every $\sim$321 d with a high eccentricity of $\sim$0.8. We monitored \hess\ with
the stereoscopic MAGIC telescopes from 2010 October to 2011 March and
detected significant VHE $\gamma$-ray emission during 2011 February, when
the system exhibited an X-ray outburst. We find no $\gamma$-ray signal in the
other observation periods when the system did not show increased X-ray
flux.  Thus \hess\ exhibits $\gamma$-ray variability on timescales of the
order of one to two months possibly linked to the X-ray
outburst that takes place about 100 days after the periastron passage. Furthermore our measurements provide for the first
time the $\gamma$-ray spectrum down to about 140 GeV and indicate no turnover of the spectrum at low energies. We compare the properties of
\hess\ with the similar $\gamma$-ray binary \lsi, and
discuss on the possible origin of the multi-wavelength emission of the source.

\end{abstract}

\keywords{ binaries: general --- gamma rays: general --- stars: individual (MWC~148)
--- X-rays: binaries --- X-rays: individual (\object{HESS~J0632+057}) }

\section{Introduction}

With the advent of the new generation of Imaging Atmospheric
Cherenkov Telescopes (IACTs) such as MAGIC, HESS and VERITAS, a
new source class, the $\gamma$-ray binaries, was established.
Only few members of this class are known to date. Among these
objects \lsi, LS 5039 and PSR~B1259$-$63 are regularly detected at
very high energy (VHE, $E > 100$~GeV) $\gamma$-rays.
All of these three systems show variable or even periodic
VHE $\gamma$-ray emission, and are spatially
unresolvable by the current generation of IACTs~\citep{2005A&A...442....1A,Magic06,LS5039_period_2006,MAGIC_lsi_periodic:2009ApJ...693..303A}.

\hess\ was discovered as an unidentified point-like VHE
$\gamma$-ray source but was
considered to be a $\gamma$-ray binary candidate because of its
spatial coincidence with the Be star \be
~\citep{HESS_discovery_2007A&A...469L...1A,Hinton_binary_hint_2009ApJ...690L.101H}.
The system was observed by VERITAS in VHE $\gamma$-rays from 2006
to 2009 with sparse sampling and the measurements did not yield
any $\gamma$-ray signal~\citep{VERITAS_ul_2009}. The derived flux
upper limits above 1 TeV were significantly below the previous detections,
thus suggesting that \hess\ was variable in VHE $\gamma$-rays. Since
all variable galactic VHE $\gamma$-ray sources known to date are
associated with binary systems\footnote{The Crab Nebula is
variable at GeV energies, but no confirmed TeV variability has been
measured up to now and thus it is not counted among the variable
VHE $\gamma$-ray sources.}, \hess\ was a very good binary
candidate. The here in detail presented VHE
$\gamma$-ray  detection simultaneously to the X-ray
outburst in 2011
February  was very recently announced
by MAGIC and VERITAS~\citep{ATel3153_VERITAS,ATel3161_MAGIC,MAGIC_ICRC_hessj0632,VERITAS_icrc_hessj0632}.

Measurements in soft X-rays with \emph{XMM-Newton} detected an
X-ray source (XMMU J063259.3+054801) at the position of \be
~\citep{Hinton_binary_hint_2009ApJ...690L.101H}. The X-ray emission
is well described by a hard power-law spectrum with energy spectral index
$\Gamma=1.26\pm0.04$, consistent with emission of synchrotron
radiation from VHE electrons, although a multi-temperature spectral
model  can also reasonably describe the data.
Furthermore, the X-ray source showed a variable flux, without
changing the spectral shape. A similar behavior is seen, e.g., in
the $\gamma$-ray binary \lsi\
(e.g. \citealt{MAGIC_lsi_xrayvhe:2009ApJ...706L..27A}). Later X-ray
observations with \emph{Swift}/XRT  found the source
but at a different flux level and with a
softer spectral index~\citep{Falcone_X_ray_2010ApJ...708L..52F}. 
Recently published \emph{Swift}/XRT observations from
2009 to 2011, display outbursts in the X-ray light curve
 from \hess\ with a periodicity of $P=321\pm5\textrm{
days}$~\citep{Bongiorno_X_Ray_2011}. These measurements also
provided evidence for hardness ratio changes with orbital phase.
The periodic X-ray emission is a strong evidence for \hess\ being
a $\gamma$-ray binary.
\emph{Chandra} high time resolution X-ray measurements during the
2011 February X-ray outburst have been used to search for pulsed
X-ray emission, but none was detected. Thus the
nature of the compact companion of \be\ remains
unknown~\citep{Rea_x_ray_2011}.

The region of \hess\ has also been observed at radio wavelengths.
The measurements conducted in 2008 with  the Very Large Array
(VLA) and the Giant Metrewave Radio Telescope (GMRT) at 5 and 
1.28~GHz, respectively, exhibited an unresolved radio source within the
position uncertainties of the VHE $\gamma$-ray source and the Be star \be
~\citep{Skilton_radio_2009MNRAS.399..317S}. A flux increase in the 5~GHz band from $0.19\pm0.04$ to
$0.41\pm0.04\mathrm{ mJy}$ showed the variability of the source on timescales
of at least one month. The radio data was well described by a power law
spectrum with energy spectral index $\alpha=0.6\pm0.2$ using non-simultaneous
data from 1.28~GHz and 5~Ghz. No extended
structures beyond the two arcsecond resolution were detected. During the
2011 February X-ray outburst very high resolution European Very Long
Baseline Interferometry Network (EVN) observations revealed a
point-like source coincident with the Be star MWC~148 within uncertainties, which evolved into an extended source with a projected size
of about 75 AU (assuming a 1.5 kpc distance), 30 days later~\citep{EVN_radio}. The peak of the
emission was displaced 21 AU between
runs, which is bigger than the orbit size (semi-major axis $\sim
2.4\textrm{ AU}$). The brightness
temperature of the source was above $2\times10^6$ K hinting to a non
thermal origin of the particles producing the radio emission. The
morphology, size, and displacement on AU scales were similar to
those found in the other gamma-ray binaries, supporting a similar
nature for HESS J0632+057 \citep{EVN_radio}. Further high
resolution measurements will be needed to understand possible
morphological changes in the radio structures along with the orbital
phase.

Optical radial velocity measurements were taken on \be\ to verify
if it is a member of a binary system and determine its orbital
parameters~\citep{Argona_opt_prop_2010ApJ...724..306A}. No
significant radial velocities were found at that time and
simulations yielded a lower limit on the possible period of the
system of $P>100\textrm{ days}$ compatible with the period found later in X-rays.
Finally, radial velocity measurements with the Liverpool telescope obtained from 2008 to 2011
have proven the binary nature of \hess/\be.
Fixing the orbital period to 321 days as obtained from the X-ray
meaurements by~\cite{Bongiorno_X_Ray_2011}  these measurements provide
for the first time the orbital parameters of the binary system.
The compact object orbits \be\ on a highly eccentric
($e=0.83\pm0.08$) orbit where the periastron passage occurs at
phase $\phi_{\mathrm{per}}=0.967\pm0.008$ using $T_0=\mathrm{MJD
 }54857.0$~\citep{Casares_optical_2011}.

In this paper, we present the VHE $\gamma$-ray measurements of
\hess\ by MAGIC from 2010 October to 2011 March. In particular we detect
VHE $\gamma$-rays only during an X-ray outburst in 2011 February and measure for the first
time the spectrum down to 140~GeV.

\section{Observations}

The observations of \hess\ were performed using the MAGIC telescopes on the Canary island of La
Palma ($28.75^\circ$N, $17.86^\circ$W, 2225~m a.s.l.), from where
\hess\ is observable at zenith angles above 22$^{\circ}$. The
MAGIC stereo system consists of two imaging air Cherenkov
telescopes, each with a 17~m diameter mirror. Each telescope
features a pixelized photomultiplier tube camera with a field of
view of about $3.5^{\circ}$. The observations were carried out in
stereo mode, meaning that only shower images which simultaneously trigger
both telescopes are recorded. The stereoscopic
observation mode provides a sensitivity so that a $5\sigma$ signal
above 300 GeV is deteced from a
source which exhibits 0.8\% of the Crab Nebula flux in 50 hours
effective time. The angular resolution is better than
$0.07^{\circ}$ above several hundred GeV and the energy resolution
is $16\%$. Further details on the design and
performance of the MAGIC
stereo system can be found in~\cite{MAGIC_stereo_performance}.

We observed \hess\ between 2010 October and
2011 March for a total of 10.6 hours. All observations
were carried out under moonlight conditions and at zenith
angles from 22 to $50^{\circ}$. The source was observed for several
nights in each month and each of these observation sets are seperated by
about 20 days. This strategy maximizes the possibility
to detect emission from \hess\ in case of a long orbital period
and with the system being active only during a short period of its
orbit i.e. one observation cycle. Due to bad
weather no data were recorded in 2010 November and 2011 January.

\section{Data Analysis}

The data analysis was performed with the standard MAGIC analysis
and reconstruction software (MARS). Events that triggered both
telescopes were recorded and further processed. The recorded
shower images were calibrated, cleaned and used to calculate
image parameters individually for each telescope. The energy of
each event was then estimated using look-up tables generated by
Monte Carlo (MC) simulated $\gamma$-ray events. In another step,
further parameters, e.g. the height of the shower maximum and the
impact parameter from each telescope, were calculated. The
gamma/hadron
classifications and reconstructions of the incoming
direction of the primary particles were performed
using the Random Forest (RF) method~\citep{magic:RF}. The RF
calculates the probability for each event to be of hadronic
origin and denotes this parameter as the hadronness of the event.
The signal selection uses cuts in the hadronness and in the
squared angular distance between the shower pointing direction and
the source position ($\theta^2$). The energy-dependent cut values
were determined by optimizing them on a sample of events recorded
from the Crab Nebula under the same zenith angle range and similar
epochs to \hess\ data. For the energy spectrum and flux, the
effective detector area was estimated by applying the same cuts
used on the data sample to a sample of MC-simulated $\gamma$-rays.
Finally, the spectrum was unfolded in energy, accounting for the
energy resolution and possible energy reconstruction
bias~\citep{magic:unfolding}.

The cuts used for producing the $\theta^2$-plot for the detection
were optimized on a Crab Nebula data sample to yield the best
sensitivity and have a higher energy threshold compared to the
cuts used to produce the spectrum. For the light curve and
integral flux calculations we chose a conservative energy
threshold of $E_{\mathrm{th}}=200$~ GeV, while the spectrum shows
reconstructed signals down to 136 GeV. Note that the systematic
uncertainties at the lowest energies dominate the total measurement
uncertainties. Using a higher energy threshold, 200 GeV, guarantees
smaller systematic uncertainties for the light curve and thus yields a
better comparison to measurements from other instruments.

\section{Results}

\begin{figure}[tbp]
  \centering
  \includegraphics[width=\linewidth]{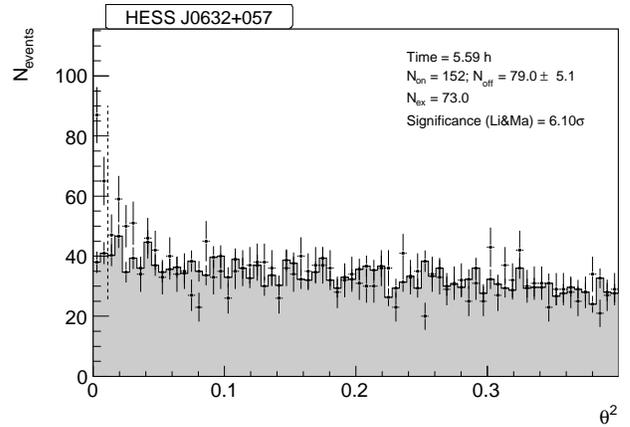}
  \caption{The squared angular distance between pointing
direction of the shower and the source position  ($\theta^2$-plot)
for the position of \hess\ (points) and the simultaneous
determined background regions (grey shaded histogram) for the
entire 2011 February MAGIC data set. The significance is calculated
according to~\cite{Li:1983fv}. $N_{\mathrm{on}}$ is the number of
events at the source position, $N_{\mathrm{off}}$ is the number of
background events, $N_{\mathrm{ex}}$ is the number of excess
events ($N_{\mathrm{ex}}=N_{\mathrm{on}}-N_{\mathrm{off}}$).
  }
  \label{fig:theta_feb}
\end{figure}

We detect VHE $\gamma$-ray emission from the \hess\ data set
recorded in 2011 February, at an orbital phase separation of $\sim$0.3  after periastron, 
with a significance of $6.1\sigma$ in 5.6 hours (see Figure~\ref{fig:theta_feb}).
The VHE $\gamma$-ray source is not
resolved by MAGIC and its extension must
be smaller than the MAGIC point spread function, whose Gaussian sigma is $0.09 \deg$ above 200 GeV.
We obtain an integral flux of $F(E>200\textrm{ GeV})=
(8.8\pm1.7_{\mathrm{stat}}\pm2.1_{\mathrm{syst}})\times 10^{-12}
\textrm{ cm}^{-2}\textrm{s}^{-1}$ which corresponds to about 4\%
of the Crab Nebula flux. Previously reported
detections~\citep{HESS_discovery_2007A&A...469L...1A} measured the
flux only above 1 TeV but agree well
within the statistical uncertainties of our measurements when
their spectrum is extended to our lower energy threshold. 
Thus we conclude that during our observations, \hess\ exhibited
similarly intense VHE
emission to the previously detected active VHE
$\gamma$-ray episodes~\citep{HESS_discovery_2007A&A...469L...1A}.

The system was only detected in the 2011 February data 
during the X-ray outburst observed by \emph{Swift}. No indication
of significant emission was found in the data from 2010 October, 2010 December or 2011
March. We denote these three months as the {\it non-detection
period} (NDP). The integration time in the individual months of the
NDP is, however,  relatively short compared to the 2011
February and we combine the NDP to have the highest possible sensitivity for a baseline VHE
flux. We obtain a flux upper limit for the NDP
of $F(E>200\textrm{ GeV}) < 3.7\times 10^{-12}\textrm{
cm}^{-2}\textrm{s}^{-1}$ at the 95\% confidence level following
the method suggested by~\cite{Rolke:2004mj}. Our flux upper limit
excludes a baseline emission down to the level of 1.7\% of the
Crab Nebula flux. No individual night during the quiescent
$\gamma$-ray state shows any indication of a signal.

\begin{figure}[tbp]
  \centering
  \includegraphics[width=\linewidth]{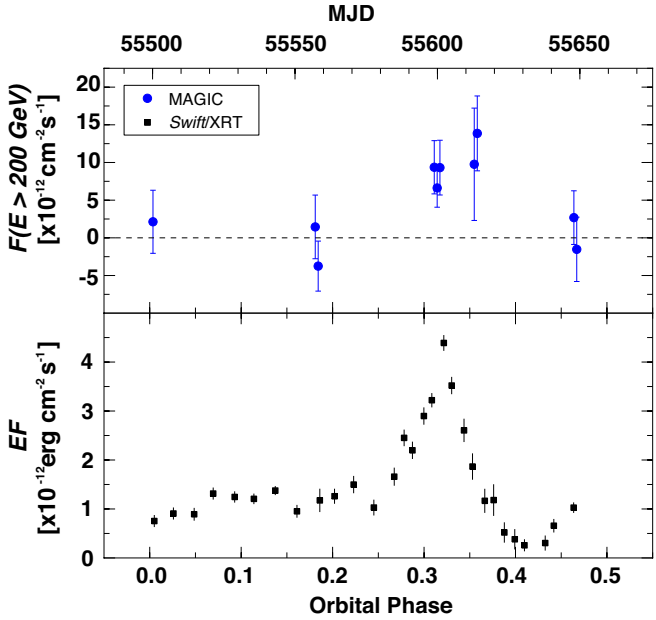}
  \caption{
The light curve of \hess\ above 200 GeV is shown in blue for the MAGIC observations
(upper panel) and  the \emph{Swift}/XRT data between 2 and 10 keV (lower panel) during the same orbital cycle as the MAGIC
  data in black. Significant emission in VHE is
  found only in 2011 February for the duration
  of about one month contemporary to the X-ray outburst.
  No variability in VHEs is seen during the active
  state. All error bars show the one-sigma
  statistical uncertainties.
  }
  \label{fig:lc}
\end{figure}

We show in Figure~\ref{fig:lc} the obtained light curve above 200 GeV of HESS 
J0632+057 for the nightly averages. The VHE gamma-ray source exhibits 
variability timescales of about one month.
Faster variability is possible but to detect it
a denser sampling of the LC is needed.
However, no short timescale (days) variability is observed during the period of
$\gamma$-ray activity in 2011 February. Under the assumption that there
is no short-time variability in the time period when no VHE
$\gamma$-ray emission is detected we conclude that the system
shows flux variations of at least a factor of two between its
quiescent and active state and that the active state must last
between 20 and 80 days in the VHE regime.

A correlation of the VHE $\gamma$-ray emission with the 2011 February 
X-ray outburst is suggestive but can not be proven statistically with our
sparsely sampled light curve. More extensive observations in VHEs are
needed for individual night correlation studies. Note that only in
the time of high X-ray activity the system was detected by MAGIC.
Whereas the X-ray light curve shows a clear peak shape for the outburst,
the VHE light curve during the $\gamma$-ray activity shows a constant flux
and no variability. Whether this constant $\gamma$-ray flux
is an artifact of the sparse sampling or a real characteristic of the
outburst cannot be determined with
these limited data. Note that a similar peak profile as in the X-ray
outburst could be present in the VHE light curve.

\begin{figure}[tbp]
  \centering
  \includegraphics[width=\linewidth]{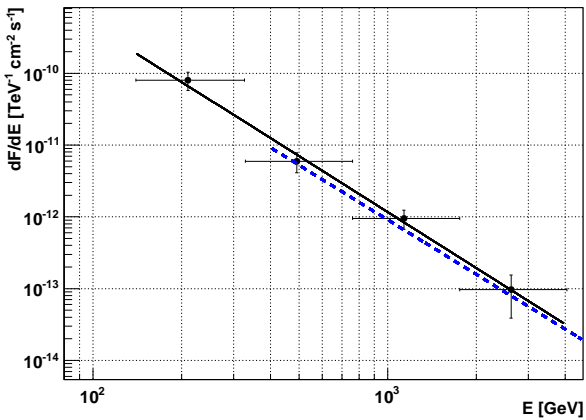}
  \caption{Differential energy spectrum of \hess\ between 136~GeV and 4
  TeV. The horizontal error bars represent the bin width whereas the
  vertical ones show the one-sigma statistical uncertainty.
  The spectrum is well described by the fitted simple power law
  with $\Gamma=-2.6\pm0.3_\text{stat}\pm0.2_\text{syst}$ shown as
  the black line. The dashed (blue) line is the spectrum obtained by
  H.E.S.S. between 2004 March and 2006 March, and is in good
  agreement with our measurement.
  }
  \label{fig:spectrum}
\end{figure}

We obtained a spectrum from the 2011 February data set and it is
compatible with a simple power law (see Figure~\ref{fig:spectrum})
with photon spectral index $2.6\pm0.3_\text{stat}\pm0.2_\text{syst}$ and
normalization $(1.2\pm0.3_\text{stat}\pm0.2_\text{syst}) \cdot 10^{-12}{\mathrm{TeV^{-1}}}\,\mathrm{cm}^{-2}\,\mathrm{s^{-1}}$.
The measured photon index is in very good agreement with the one
previously published by H.E.S.S. ($\Gamma=2.53\pm0.26_\text{stat}\pm0.2_\text{syst}$), although their spectrum was obtained
at energies above 400 GeV~\citep{HESS_discovery_2007A&A...469L...1A}. No indication of a turnover is found in
the newly opened energy range by MAGIC.

\section{Discussion}
The  VHE gamma-ray data obtained during the periodic X-ray outburst of
HESS J0632+057 that took place during 2011 February suggests that the VHE light curve shows similar outbursts like
the X-ray light curve.  The detection of VHE $\gamma$-ray 
emission only during the X-ray outburst suggests a common origin 
although our data are too sparsely sampled to allow night to night 
correlation studies.
The MAGIC detection of the source  during the peak
of the X-ray outburst yielded a similar flux level and energy
spectrum as those obtained by HESS four years before~\citep{HESS_discovery_2007A&A...469L...1A}.  Similar
spectral shape and flux levels indicate that the same processes might be at work during the $\gamma$-ray active states. In case of a periodic modulation with a period of about
321~days, such a behavior would be expected. 

For the first time we could measure the spectrum of \hess\ between
136 and 400~GeV, and find no evidence for a spectral break or a
deviation from a simple power law. This is a common feature found in
other binaries which exhibit outbursts (e.g. \lsi~\citealt{MAGIC_lsi_periodic:2009ApJ...693..303A}). Thus the
 turnover in the spectrum must lie below the energy threshold of our
 observations.  All currently known $\gamma$-ray binaries
show their maximum emission in the high
MeV to GeV energy range. This might be the case for \hess\
although it has not yet been detected by \emph{Fermi}/LAT. 
The integration time required to detect \hess\ with \emph{Fermi}/LAT will depend
strongly on the source's duty cycle and the spectral properties at MeV
to GeV energies.

In several recent publications the similarity between \hess\ and \lsi\ was stressed
based on their similar multiwavelength
emission~(e.g.~\citealt{Hinton_binary_hint_2009ApJ...690L.101H},~\citealt{Skilton_radio_2009MNRAS.399..317S}). We note that there
might be some differences in the VHE emission. 
\lsi\ shows complex VHE behavior such as variability on timescales as short as one day
(e.g. \citealt{MAGIC_lsi_xrayvhe:2009ApJ...706L..27A}), and
different VHE $\gamma$-ray flux
states~\citep{VERITAS_LSI_stateS_2011ApJ...738....3A,lsi_flux_states}. Revealing
similar behavior in 
\hess, requires higher sensitivity and better temporal sampled data
than available to date and thus the degree of similarity between \lsi\
and \hess\ might be smaller compared to the contemporary view.
Interestingly, the orbital phase lag for the detected VHE $\gamma$-ray
emission in \hess\ is
quite close to that in \lsi, about  0.3 after
periastron passage \citep{Li_11,Casares_optical_2011}. Similar
processes might produce both the X-ray and the VHE $\gamma$-ray radiation. However,
different  spatial- and time-scales of the emitter, given the wider and more
eccentric orbit in \hess, and different star-emitter-observer geometries
(important in the leptonic scenario), make any direct comparison
difficult.  A proper characterization of the radiation and magnetic
fields, and of possible adiabatic losses, is required. Said
that, however, we can interpret our \hess\ data in the context of the X-ray
data, as done for \lsi\ (see \citealt{MAGIC_lsi_xrayvhe:2009ApJ...706L..27A}). 

In the case of \lsi, the X-ray/VHE emission was suggested to originate from a homogeneous
leptonic emitter relatively close to the compact object under dominant IC
losses. This scenario, already explored
in~\cite{Hinton_binary_hint_2009ApJ...690L.101H}, is supported by the similar fluxes in X-rays and
VHE, which is hard to explain in the context of an homogeneous emitter producing
$\gamma$-rays (and the X-ray emitting $e^{\pm}$-pairs) via proton-proton
collisions. In the case of \hess, however, the X-ray luminosity is slightly
lower than the VHE luminosity. Although in this case the fluxes and spectra
still allow the leptonic homogeneous scenario with dominant IC losses, a
hadronic homogeneous emitter cannot be discarded. Therefore the question of
the hadronic or leptonic nature of the emitter in \hess\ remains unsolved.

To distinguish between the hadronic and leptonic pictures, and the one
population hypothesis for the VHE $\gamma$-ray and the X-ray emission
needs a better sampling of the light curve at the relevant orbital
phases. A spectral index correlation study would shed light on the
origin of the emission. Such a study will require long and frequent
individual night observations to provide high enough sensitivity for
more precise spectral information. Fortunately, due to the X-ray periodicity, such
measurements can be planned well in advance. Complementary information from the
GeV band will also be valuable, e.g. revealing the spectral shape in the
currently unmeasured energy range below 100 GeV and above 100 MeV
since this information could reveal the population(s),
of particles involved in the emission processes.

\section*{Acknowledgments}

We would like to thank the Instituto de Astrof\'{\i}sica de
Canarias for the excellent working conditions at the Observatorio
del Roque de los Muchachos in La Palma. The support of the German
BMBF and MPG, the Italian INFN, the Swiss National Fund SNF, and
the Spanish MICINN is gratefully acknowledged. This work was also
supported by the Marie Curie program, by the CPAN CSD2007-00042
and MultiDark CSD2009-00064 projects of the Spanish
Consolider-Ingenio 2010 programme, by grant DO02-353 of the
Bulgarian NSF, by grant 127740 of the Academy of Finland, by the
YIP of the Helmholtz Gemeinschaft, by the DFG Cluster of
Excellence ``Origin and Structure of the Universe'', and by the
Polish MNiSzW Grant N N203 390834.

Facilities: \facility{MAGIC}

\bibliographystyle{astron}

\end{document}